\documentclass[fleqn,12pt,twoside]{article}
\usepackage{espcrc1}

\usepackage{graphicx}
\newcommand{\about}{$\simeq$}

\def\degree{\relax\ifmmode^\circ \else $^\circ$\fi}

\newcommand{\Al}{$^{26}$Al\ }
\newcommand{\Fe}{$^{60}$Fe\ }
\newcommand{\Na}{$^{22}$Na\ }
\newcommand{\Ni}{$^{56}$Ni\ }
\newcommand{\Ti}{$^{44}$Ti\ }

\title{Gamma-ray Line Astronomy
\footnote{\it accepted for Nucl.Phys.A (Proc. NIC8, Vancouver 2004)}
}

\author{R. Diehl\address[MPE]{Max-Planck-Institut f\"ur extraterrestrische Physik,
              D-85741 Garching, Germany}
        \thanks{INTEGRAL/SPI results reported here have been obtained through collaborative efforts of the team of
SPI scientists, thanks to the excellent performance which resulted from the dedication of
SPI detector and subsystem experts, ESA mission operations and the Russian launch vehicle.
SPI has been completed under the responsibility and leadership of CNES.
We are grateful to ASI, CEA, CNES, DLR, ESA, INTA, NASA and OSTC for support.
}
}%
\begin{document}
\maketitle
\begin{abstract}
Gamma-ray lines from radioactive isotopes, ejected into interstellar space
by cosmic nucleosynthesis events, are observed with new space telescopes.
The Compton Observatory had provided a sky survey for the isotopes
$^{56}$Co, $^{22}$Na, $^{44}$Ti, and $^{26}$Al, detecting supernova radioactivity
and the diffuse glow of long-lived radioactivity from massive stars in the Galaxy.
High-resolution spectroscopy 
is now being exploited with Ge detectors: Since 2002, with ESA's INTEGRAL
satellite and the RHESSI solar imager two space-based Ge-gamma-ray 
telescopes are in operation, measuring Doppler
broadenings and line shape details of cosmic gamma-ray lines.  
First year's results include a detection and line shape measurement of annihilation 
emission, and $^{26}$Al emission from the inner Galaxy and from the Cygnus region.
$^{60}$Fe gamma-ray intensity is surprisingly low; it may have been detected by RHESSI
at 10\% of the \Al brightness, yet is not seen by INTEGRAL. \Ti emission from Cas A and
SN1987A is being studied; no other candidate young supernova remnants have been found through $^{44}$Ti. 
 \Na from novae still is not seen.
\end{abstract}

\section{INTRODUCTION}

Radioactive isotopes as common by-products of nucleosynthesis in cosmic 
sources can be studied through their gamma-ray emission \cite{dieh98}. 
Candidate sources are supernovae and novae, but also the winds from massive
stars. 
Continuum emission from Comptonized or thermalized radioactive energy
is also exploited as a measurement of nucleosynthesis, but less direct:
Conversion of original radioactive
energy must be calculated through complex radiation transport models. 
Alternatively, characterisic
isotopic samples of the nuclosynthesis source, as conserved in circum-source
dust grains, allow significant nucleosynthesis 
inferences, in particular on AGB stars, but more recently even on supernovae
and novae. But here the physics of dust condenstation near the production site,
and transport and processing history of such grains is uncertain.
Other measurements of cosmic nucleosynthesis are also rather indirect,  
e.g. the 
analysis of X-ray line emission from ionized-gas portions of supernova 
remnants and galactic-halo gas. 
Radioactive decay in interstellar space is unaffected by 
physical conditions in/around the source such as temperature or density, and
decay gamma-rays are not attenuated along the line-of-sight due to their penetrating nature
(attenuation length \about~few g~cm$^{-2}$). 
Therefore, such decay gamma-rays with satellite-borne 
telescopes provide the most direct measurement of
the existence of freshly-produced isotopes.


Various gamma-ray telescope experiments have established this new window for the study
of cosmic nucleosynthesis during the past three decades \cite{dieh98}:
\begin{itemize}
\item{}
Interstellar \Al has been mapped along the plane of the Galaxy,
confirming that nucleosynthesis is an ongoing process 
\cite{dieh95,pran96,plue01}.
\item{}
Characteristic Ni decay gamma-rays have been observed from SN1987A
\cite{tuel90,matz88,kurf92}, directly
confirming supernova production of fresh isotopes up to iron group nuclei.
\item{}
\Ti gamma-rays have been discovered \cite{iyud94,scho00}
from the young supernova remnant Cas A,
confirming models of $\alpha$-rich freeze-out for core-collapse supernovae.
\item{}
A diffuse glow of positron annihilation gamma-rays has been recognized
from the direction of the inner Galaxy \cite{purc97,kinz01}, 
consistent with nucleosynthetic
production of $\beta^+$-decaying radioactive isotopes 
\end{itemize}

These  results have led to new questions:
\begin{itemize}
\item{}
What fraction of radioactive energy is converted into other forms
of energy in supernovae? (This addresses: absolute normalization 
of indirectly-inferred radioactive amounts from \Ni in SNIa, and
\Ti in core-collapse SNe; positron leakage from 
supernovae; morphology of expanding
supernova envelopes, "bullets", filaments, jets). 
\item{}
How good are our (basically one-dimensional) models for nova and supernova
nucleosynthesis, in view of important 3D effects such as rotation and
convective mixing? (This addresses:  \Ti mass
ejected from regions near the mass cut between compact remnant and 
ejected supernova envelope; nova \Na yields and the 
seed compositions for explosive hydrogen burning in novae).
\item{}
What is the range of physical conditions expected for nucleosynthesis
events? (This addresses:  stellar mass distributions and
supernova rates in massive-star clusters; clustering of events in space
and time; self-enrichment; triggered
star formation in dense, active nucleosynthesis regions; first stellar generations when 
metallicity was extremely low).
\item{}
How are ejecta and energy from nucleosynthesis events fed back into the
interstellar medium? (This addresses:  scales and time constants of chemical
evolution of interstellar gas; morphology of the interstellar medium; 
spatial pattern of star formation; nucleation of dust and its processing 
by interstellar shocks; acceleration of cosmic rays) 
\end{itemize}

With INTEGRAL and RHESSI, 
now a new step is taken, 
improving sensitivities by \about~an order of magnitude: 
The spectrometer SPI \cite{vedr03,roqu03} on INTEGRAL \cite{wink03i} 
is based on 19 hexagonal Ge detector modules, 
each one 7~cm thick and 5.5~cm wide (flat-to-flat across their hexagonal
shape), arranged in a densely packed detector plane of 500~cm$^2$ total area. 
Failure of two of the detector modules 
(\#2 in Dec 2003, \#17 in July 2004) has reduced the effective area now by \about~10\%.
The Ge detectors are illuminated through a coded mask made of  3~cm of tungsten,
defining a telescope field-of-view of 16$^\circ$.
Detector temperature is maintained at \about~90~K by Stirling coolers.
The accumulated damage from cosmic-ray bombardement in the detector crystals
leads to \about~10\% degradation of spectral resolution and an asymmetric
response. This is cured every \about~6~months by heating up the camera to 
\about~105$^\circ$~C for \about~1.5 days ("annealing"), a procedure which has been 
exercised successfully in orbit four times already, re-establishing 
the original spectral response and resolution of \about~3~keV at 2~MeV.
Complementing this, the RHESSI solar observatory satellite \cite{lin02} 
with its imaging spectrometer instrument 
carries similar fine-spectroscopy Ge crystals.
Although pointed at the Sun, this has proven sensitive enough for 
diffuse cosmic gamma-ray emission studies, when using the Earth as a
shadowing device.
With these instruments, we now can observe
kinematic signatures from Doppler-shifted energy values in 
expanding/accelerated radioactive material from sources
of cosmic nucleosynthesis. 

\begin{figure}
\centering
\includegraphics[width=6.4cm]{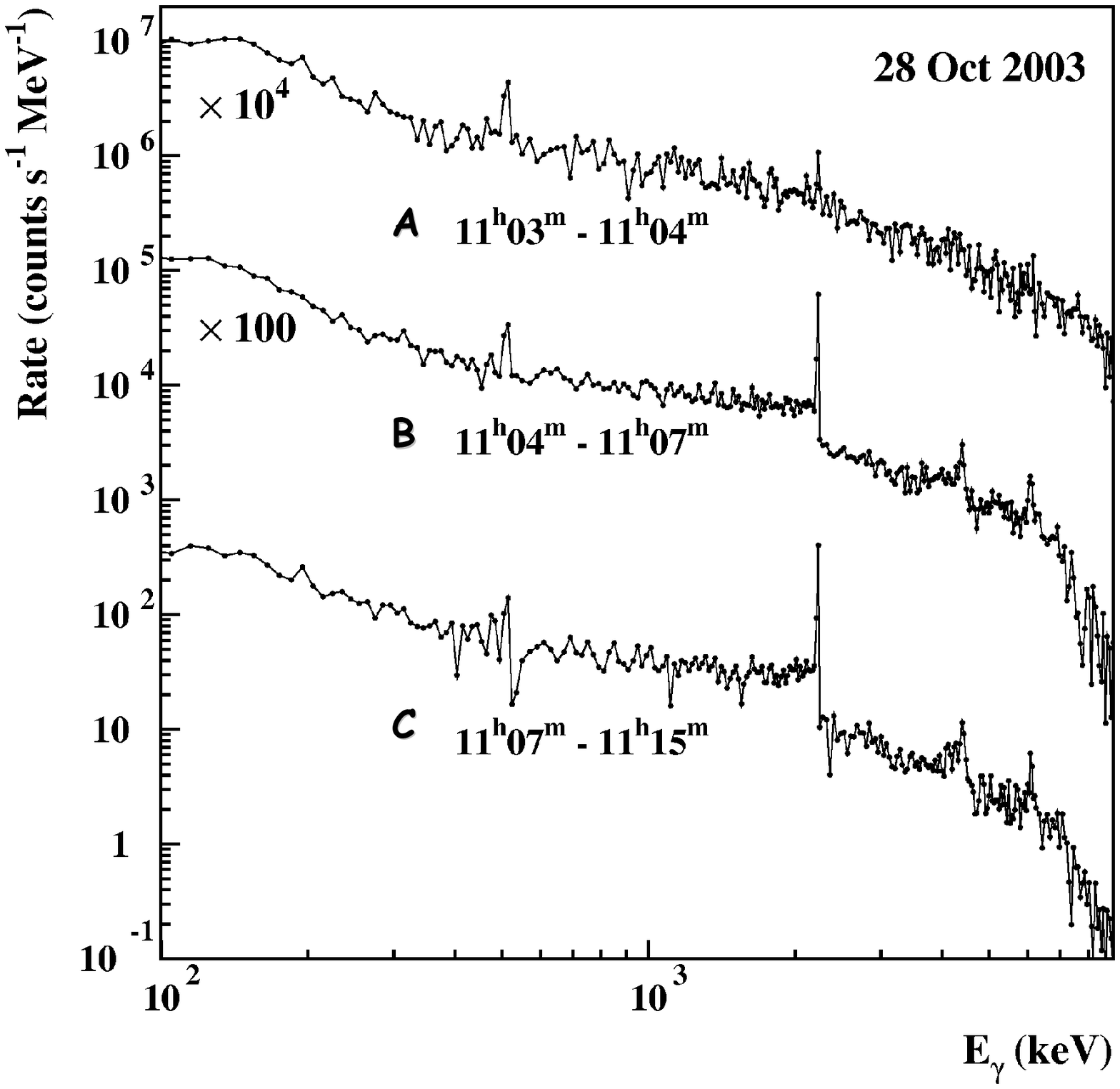}
\includegraphics[width=7.6cm]{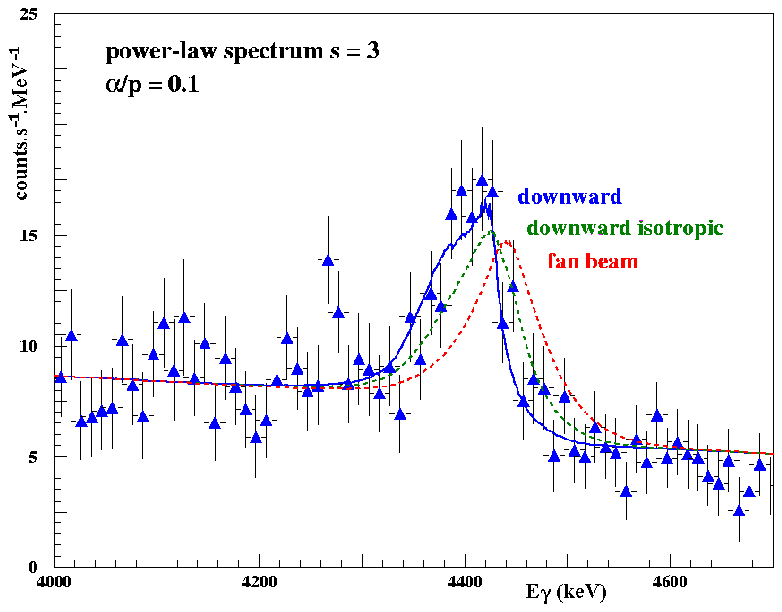}
\caption{SPI measurement of the pectral variations over 
three successive phases of the Oct 28
solar flare (left). The $^{12}$C line shape from 
late flare phases (B,C) is compared with models 
for different geometries of the accelerated
particle beam (right) (see \cite{gros04}).  
}
\label{fig_solarflare}
\end{figure}

\section{RESULTS}
In their first two mission years, INTEGRAL and RHESSI results have already demonstrated
the useful complement of such measurements for the field:
\begin{itemize}
\item{}
Solar flare observations allowed measurements of flare spectra and in particular excited-particle flow
geometries \cite{lin03,smit04,gros04}.
\item{}
Gamma-ray line emission from \Al and positron annihilation have been
measured, imaging their sources in the Galaxy, and determination
of source velocity characteristics is underway \cite{dieh03,knoe04c,jean03,gues04}.
\item{}
Search for $^{60}$Fe emission has been partly successful, with a marginal RHESSI
result and new limits from INTEGRAL \cite{smit03,smit04,knoe04f}.
\item{}
Search for \Ti emission from Cas A and SN1987A, and from unknown but expected
sources in the inner Galaxy is underway \cite{rena04,kien04}.
\end{itemize}

\subsection{Solar Flares}
Gamma-ray spectroscopy of solar flares is a prime diagnostic of
particle acceleration mechanisms, the Sun a unique nearby laboratory.
Ions are energized in the solar magnetosphere and collide with matter of the upper 
solar atmosphere. Collisionally-excited nuclei de-excite, emitting characteristic
gamma-ray lines. 
The physical process and energy source of particle acceleration
are essentially unknown; their study is the main objective of the
RHESSI space experiment.
Imaging the footpoints of gamma-ray emission with respect to to the 
flare loops, interesting offsets for neutron capture versus 
de-excitation lines have been revealed \cite{lin03,smit04}. 
For one of the stronger recent flares,
also INTEGRAL's spectrometer measured C and O line shapes, even improving on RHESSI's spectroscopic
precision. Observed line redshifts are rather well determined
and consistent with downward beaming of energetic particles. 
Here, the beam geometry
could be constrained to a narrow downward-directed one, 
rather than isotropic or fan-like \cite{gros04} (see Fig.\ref{fig_solarflare}).

\begin{figure}
\centering
\includegraphics[width=7.5cm]{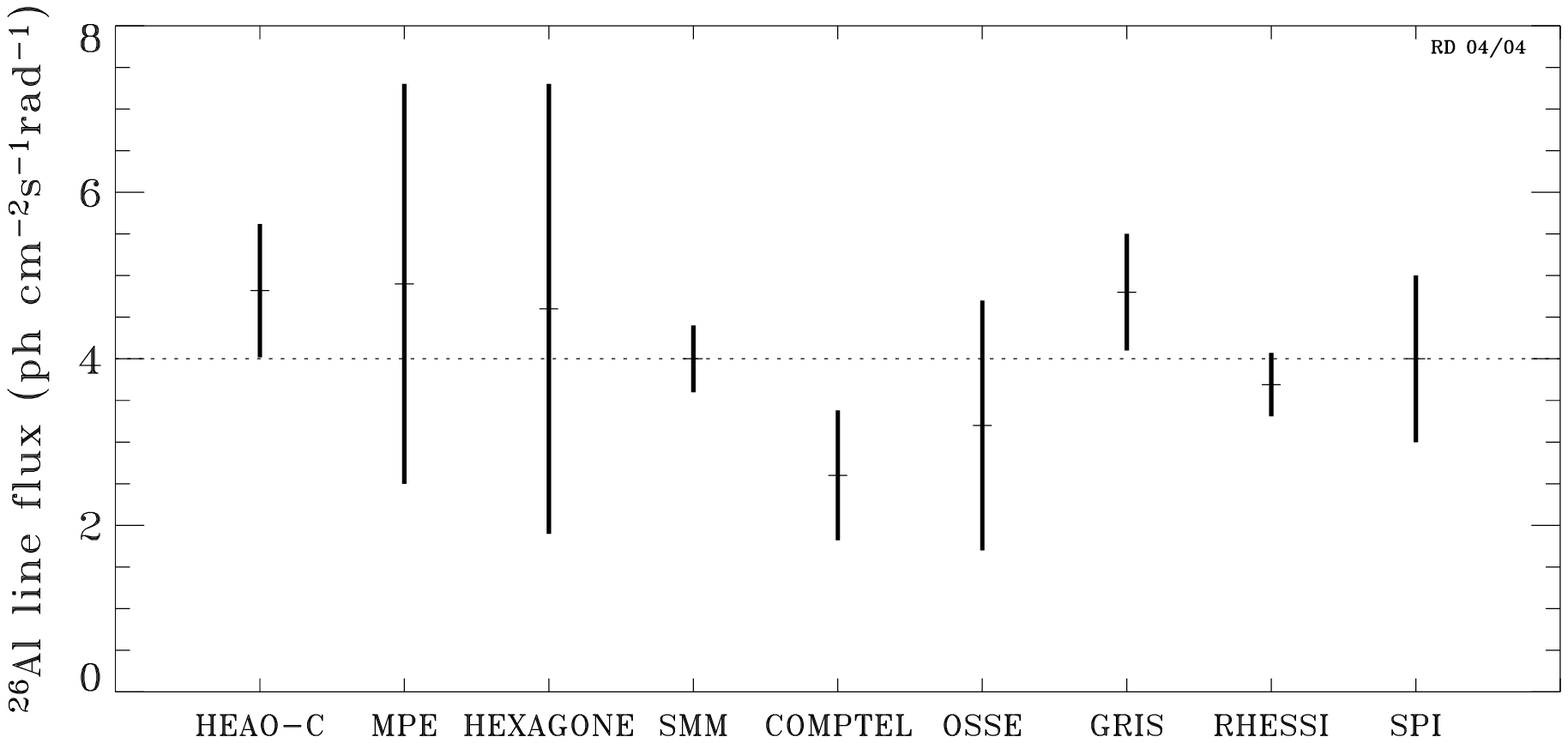}
\includegraphics[width=7.5cm]{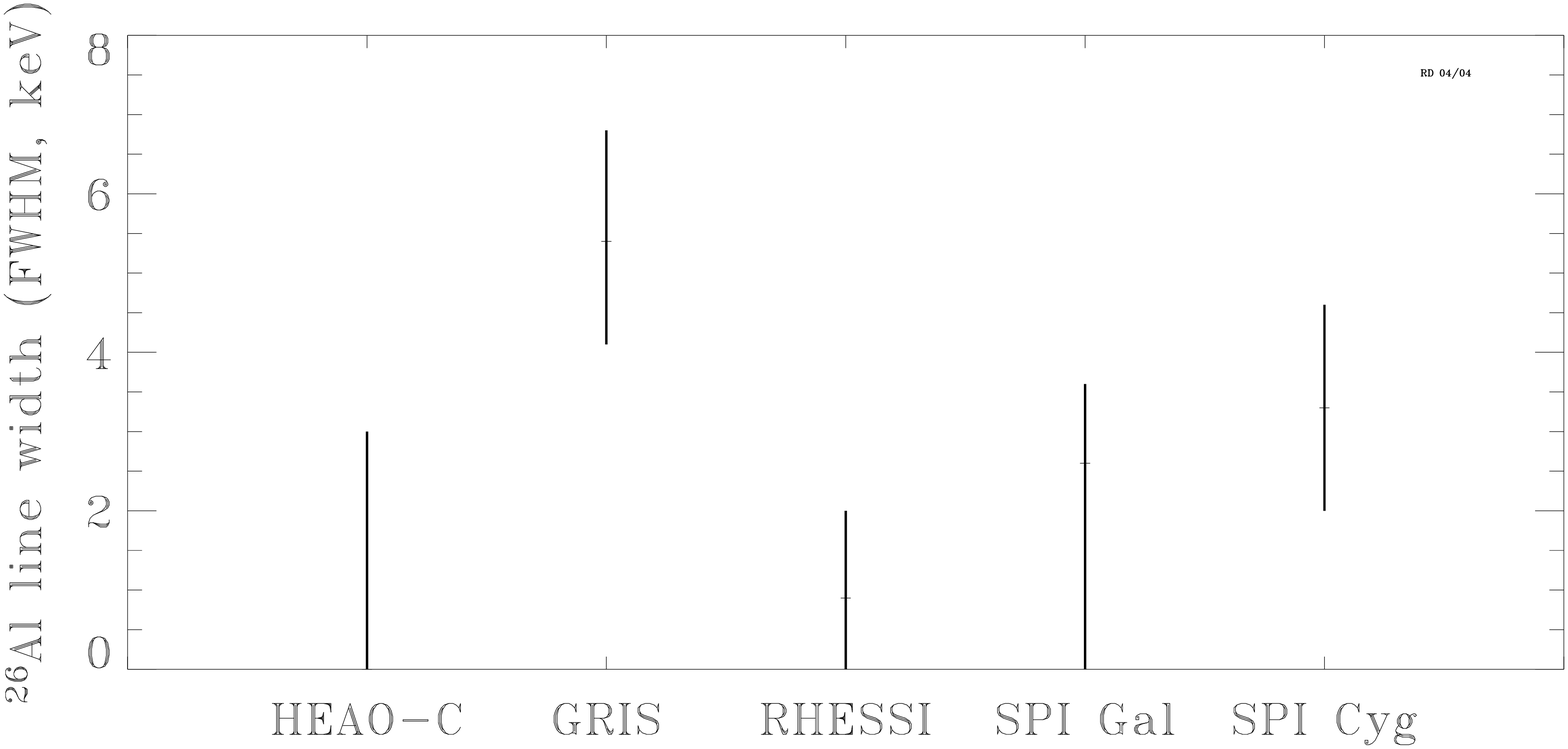}
\caption{Intensity measurements (left) and line width measurements (right)
from different experiments for \Al emission
from the inner Galaxy, and for Cygnus (rightmost datapoint). }
\label{diehl_26Al_experiments}
\end{figure}

\subsection{Interstellar \Al and \Fe}

\begin{figure}
\centering
\includegraphics[width=7cm]{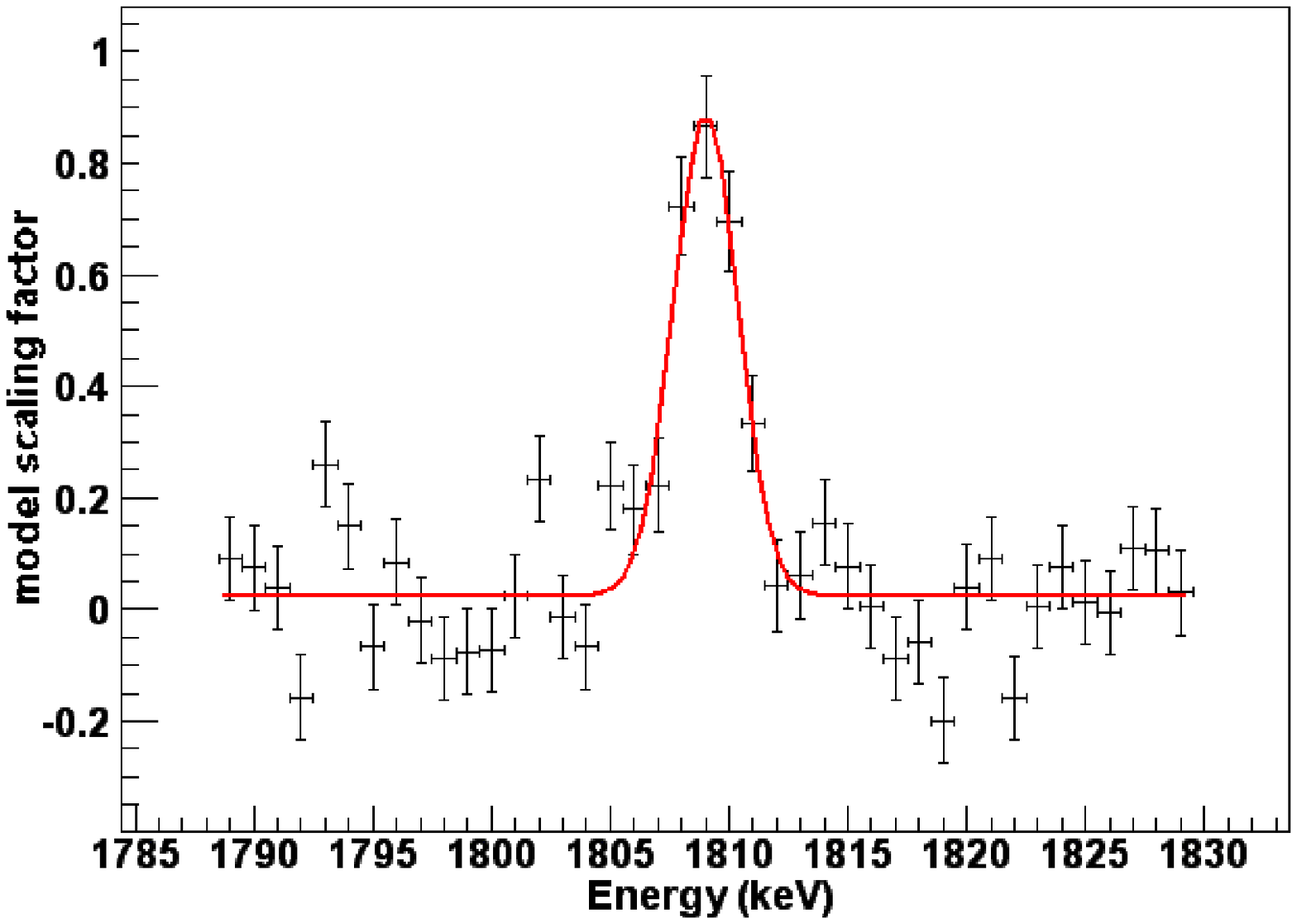}
\includegraphics[width=7cm]{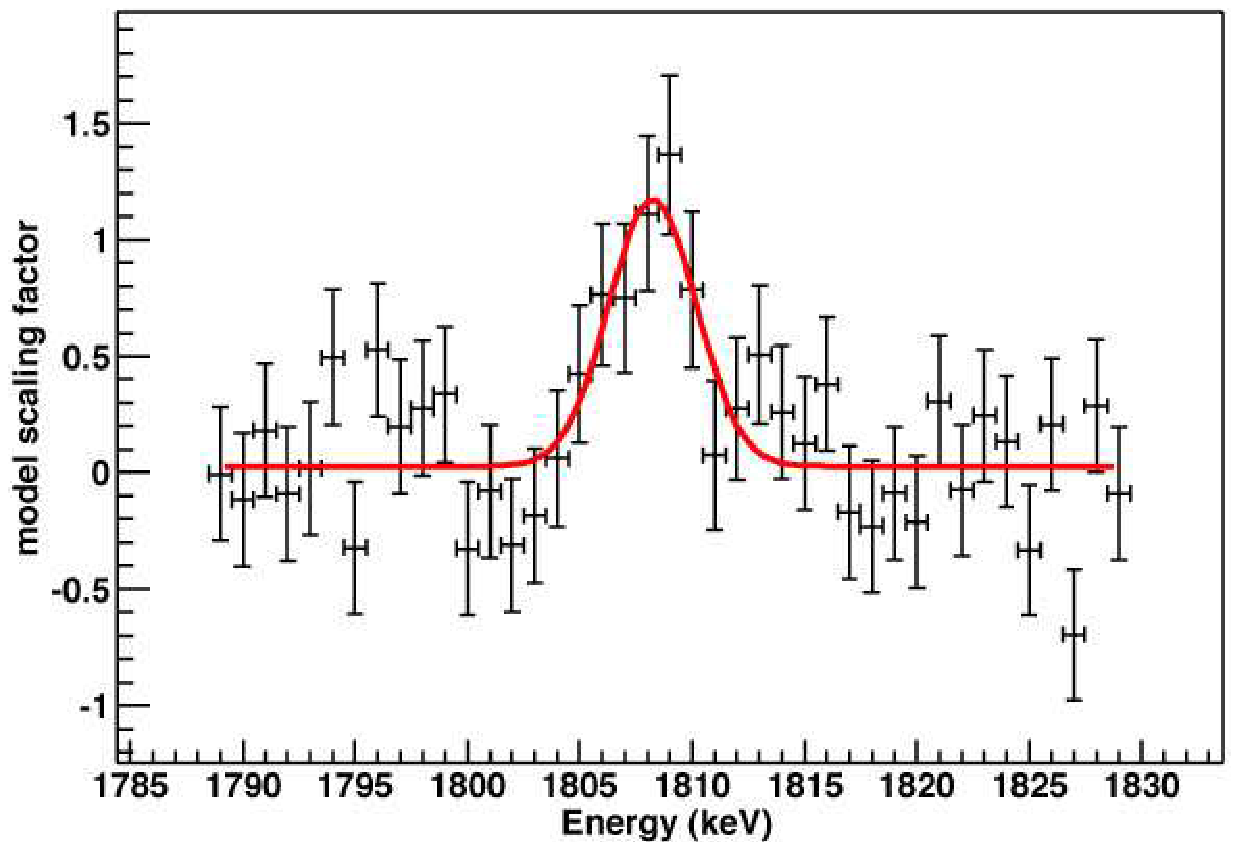}
\caption{SPI \Al line spectra for the inner Galaxy (left) and the
Cygnus region (right). The spectra are derived by fitting skymap intensities
per energy bin. Exposures are \about~4Msec each, from the Core Program
inner-Galaxy survey, and from 
Cygnus region calibrations and Open-Program data.  
}
\label{fig_spec_obsfit}
\end{figure}

Precision follow-up measurements of 1808.7 keV emission from 
Galactic \Al have been one 
of the main science goals of the INTEGRAL mission \cite{wink03i},
after the COMPTEL sky survey 
\cite{plue01,knoe_img99,ober97,dieh95} had
mapped structured \Al emission, extended along the plane of the Galaxy.
Models of \Al emission from the Galaxy and specific localized source regions
have been based on knowledge about the massive-star populations, and suggest that
such stars dominate \Al production in the Galaxy 
\cite{pran96,knoe_mod99,knoe_phd99}. 
Galactic rotation and dynamics of the \Al gas ejected into the 
interstellar medium are expected to leave
characteristic imprints on the \Al line shape  \cite{kret03}. 
In particular after the GRIS balloon experiment reported 
a significantly-broadened line \cite{naya96}, which translates into a Doppler
broadening of 540~km~s$^{-1}$, 
alternative measurements of the \Al line 
shape have been of great interest. 
 Considering the $1.04 \times 10^6$~y decay time of \Al
such a large line width is hard to understand, and requires
either kpc-sized cavities around \Al sources or major \Al 
condensations on grains \cite{chen97,stur99}. 

Current results on large-scale
Galactic \Al line flux and width measurements are summarized 
in Fig. \ref{diehl_26Al_experiments}.
From INTEGRAL/SPI spectral analysis of a subset of the first-year's inner-Galaxy deep
exposure ("GCDE"), \Al emission was clearly detected  (Fig. \ref{fig_spec_obsfit} left) 
\cite{dieh03,dieh04} (5--7$\sigma$; through fitting 
of adopted models for the \Al 
skymap to all SPI event types over an \about~80~keV energy range around the \Al line).  
The line width was found consistent with SPI's instrumental
resolution of 3~keV (FWHM). 
Therefore, these early SPI results 
support RHESSI's recent finding \cite{smit03}
that the broad line reported by GRIS \cite{naya96} probably cannot be confirmed
(Fig. \ref{diehl_26Al_experiments}).
On the other hand, the first spectrum generated from SPI data for the Cygnus region 
(see Fig. \ref{fig_spec_obsfit} right for single-detector events \cite{knoe04c}) suggests
that the line may be modestly broadened for this region. This
may be caused by locally-increased interstellar turbulence in the particularly
young stellar associations of Cygnus.

\begin{figure}[ht]
\centering   
\includegraphics[width=9cm]{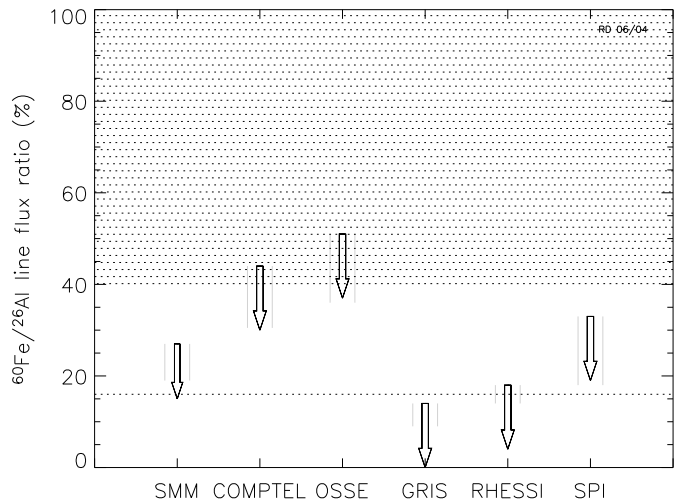}
\caption{Limits on the \Fe to \Al gamma-ray brightness ratio,
from several experiments fall below current expectations: 
An often-cited theoretical value from the 
Santa Cruz group \cite{timm96} is indicated as dotted
line, the dotted area marks the regime suggested by 
recently-updated models, from \cite{pran04}.
	 }
      \label{diehl_fe60limits}
\end{figure}

The fact that \Fe has not been clearly seen from the same source regions
appears surprising \cite{pran04}, given that $^{26}$Al-ejecting 
massive stars are expected to also eject \Fe in 
substantial amounts \cite{timm96,limo04}. RHESSI reported a
marginal signal \cite{smit04} at the 10\%-level of \Al brightness, 
and also SPI's upper limit is substantially below the \Al brightness \cite{knoe04f}.
If massive stars are the dominating \Al sources, recent models of nucleosynthesis
(see \cite{pran04,limo04})
even increase \Fe production in late evolutionary phases near the
supernova at the bottom of the He shell, or even more efficiently in the C shell.
This is mainly due to increased neutron capture cross sections for Fe isotopes, 
and a reduced $^{59}$Fe $\beta$-decay rate. 
The now-expected ratio falls in a range between 40 and 120\%, 
depending on how the interpolation between the few calculated stellar-mass
gridpoints is done, and which of the models are taken as baseline; in comparison,
the IMF choice appears uncritical. In any case, revised expectations clearly
lie above the experimental limits for Galactic \Fe emission (Fig.~\ref{diehl_fe60limits}); the answer
may be found in model detail, possibly
in nuclear-physics issues related to the (uncertain) Fe neutron capture reactions.

\subsection{Positron Annihilation}

\begin{figure}[ht]
\centering   
\includegraphics[width=7.2cm]{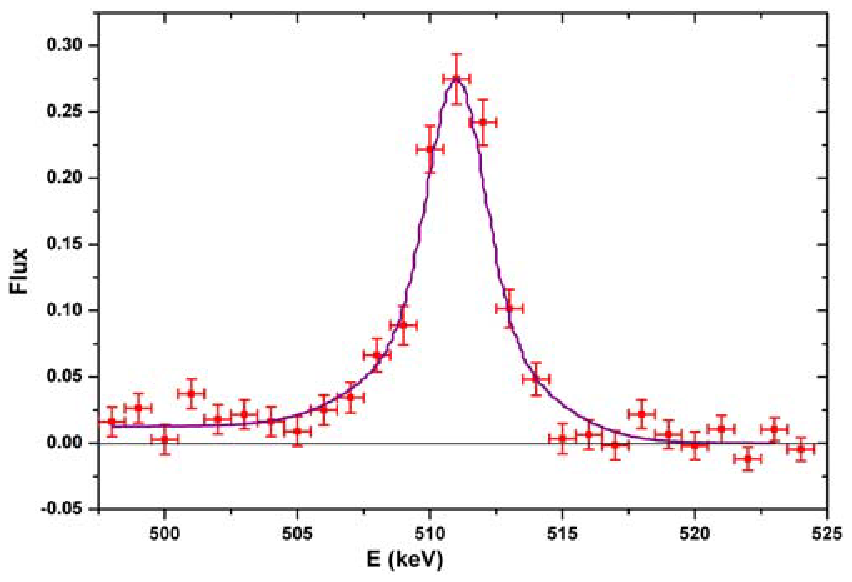}
\includegraphics[width=7.6cm]{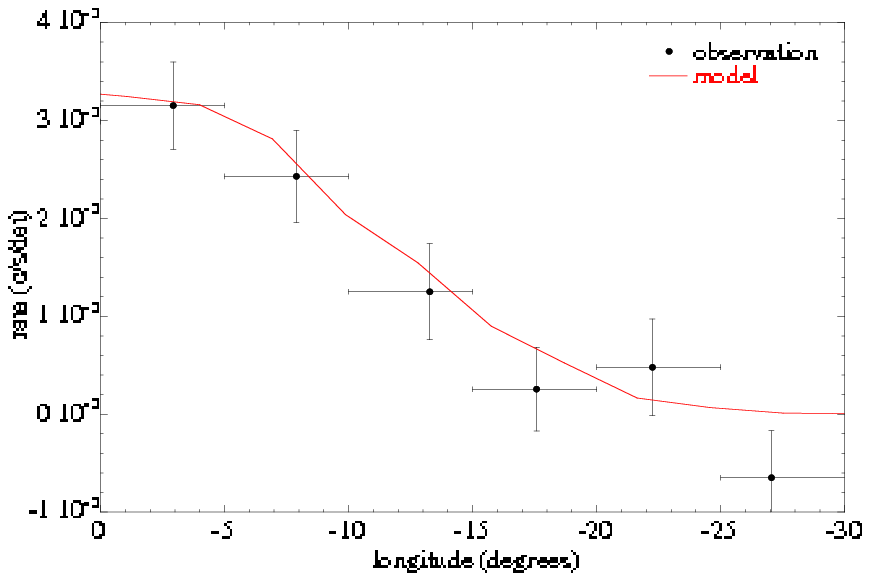}
\caption{SPI measurement of the positron annihilation gamma-ray line from the 
             inner Galaxy.
			 The line is confirmed to be instrinsically broadened \cite{lonj04} (left: 
			 intensity in units 10$^{-4}$ph~cm$^{-2}$s$^{-1}$ for
			  1~keV wide bins, fitted with a composite model of gas and dust, from \cite{gues04}). 
			 The positron annihilation gamma-ray line intensity in the inner Galaxy 
   as observed with SPI, compared to the longitude distribution expected
   from a Gaussian-distributed model \cite{jean03} (right).
	 }
      \label{diehl_511}
\end{figure}
Positrons (e$^+$) are produced upon $\beta^+$-decay of proton-rich unstable isotopes.
Other sources  are expected to contribute to the e$^+$ budget, however: 
Plasma jets ejected from 
pulsars \cite{zhur87,kenn84} or microquasars \cite{mira92} will
produce e$^-$e$^+$ beams
as a consequence of rotational magnetosphere discharges or accretion, 
respectively, and annihilation of dark-matter particles attracted by 
the gravitational potential of the Galaxy may produce distributed 
e$^-$e$^+$~pairs \cite{boeh03}.
The fractional contribution from nucleosynthesis sources to the e$^+$
budget within the inner Galaxy regions is estimated to range from \about~30\% to
100\% (mainly SNIa and novae  through 
$^{56}$Co, $^{13}$N, and $^{18}$F). 
Substantial uncertainty in e$^+$ yields arises from the unknown e$^+$ escape fractions 
from these sources. 
A lower limit is placed from observed \Al, at \about~3~10$^{-4}~$ph~cm$^{-2}$s$^{-1}$.

The intensity in the annihilation line is 
$9.9 ^{+4.7}_{-2.2} \times 10^{-4}$~ph~cm$^{-2}$s$^{-1}$, as derived from first
analysis of SPI data from a part of the inner-Galaxy deep exposure (GCDE)
of the first mission year \cite{lonj04,jean03} 
(see Fig.\ref{diehl_511}), which is consistent with
previous measurements and theoretical predictions.
This corresponds to a positron production
rate in the inner Galaxy on the order of 10$^{43}$s$^{-1}$
for an assumed steady state \cite{rama94}.

Positrons with high energies have a low probability for annihilation.
Generated at typical energies of \about~MeV, only a small fraction will
annihilate before thermalization along their trajectories in the ISM.
Once thermalized, positrons annihilate
preferentially through the intermediate formation of positronium atoms,
or on the surface of charged dust particles. 
This leads to different gamma-ray spectral components from each of
these annihilation channels, a composite shape of the 511~keV line
and an underlying Ps continuum \cite{gues91}. 
Detailed 511~keV line shape measurements are an important diagnostic, (a) because its
measurement is easier with high spectral-resolution experiments than 
constraining the spectrally-distributed and background-contaminated continuum,
and (b) because the details of the annihilation line shape will encode 
kinematics and thermal properties of the annihilation regions. 
As a complication, the lifetime of \about~MeV positrons  
in interstellar space can be substantial \cite{gues91}, up to 10$^5$~y, 
so that positrons  may travel significant (few 100 pc) distances between 
their sources and the locations of their annihilation. 

SPI detects Ps continuum \cite{stro04} and 511 keV line \cite{jean03}, 
and finds the line to be significantly broadened, from deconvolution with
the instrumental resolution after subtraction of the (strong) instrumental
background line \cite{lonj04}.  
The SPI value of $2.95 (\pm 0.6)$~keV (FWHM) is on the high side of values
measured by previous instruments
(HEAO-C \cite{maho94}: 1.6 $\pm$ 1.3; 
GRIS \cite{leve93}: 2.5 $\pm$ 0.4; TGRS \cite{harr98}: 1.81 $\pm$ 0.54). 
First composite source process models 
suggest that annihilation on grains are
a significant contribution, in order to obtain a sufficiently narrow line profile
satisfying the SPI measurement \cite{gues04}.

A diffuse nature of the source is expected both from  
radioactive and from dark matter sources, while localized
emission / hot spots would be expected if annihilation near compact sources
(microquasars, pulsars) is significant. Therefore, models to interpret 
previous measurements with instruments
of rather modest (few degrees) imaging resolution have been composed
from disk contributions (diffuse radioactivity from bulge and disk novae
and other sources, or latitudinally more-extended warm or hot ISM gas
models) and from point sources for candidate e$^-$e$^+$ producers
such as 1E~1740.7-2942. 

From OSSE scans of the inner Galaxy with its field-of-view of 11.4~x~3.8$^\circ$, the
spatial distribution was found to be best represented by a Gaussian with
an extent of \about~5$^\circ$~(FWHM) in longitude and latitude \cite{kinz01,miln01}.
Surprisingly, only a minor disk-like component was observed, and
the "bulge" component appeared rather extended; furthermore, there
was indication of asymmetry \cite{purc97}, with an excess of annihilation emission
in the northern hemisphere towards latitudes of \about~10$^o$.
Yet, no mapping of annihilation emission was available
along the disk of the Galaxy outside this inner region;
this is one of the current projects for the SPI instrument on INTEGRAL.

SPI data \cite{jean03,knoe03} confirm OSSE's 
findings: The 511~keV line flux, as it varies for different exposures successively pointed
along the plane of the Galaxy 
(Fig. \ref{diehl_511}) suggests that
diffuse emission extends over a large volume. A Gaussian spatial distribution is 
constrained to an extent (FWHM) of
6$^\circ$--18$^\circ$ \cite{jean03,weid04}. First imaging attempts do not show signs of 
asymmetry, but confirm the extended and rather smoothly-distributed emission, 
both in longitude and latitude \cite{knoe03}. There is no hint for emission
from the disk of the Galaxy yet; such emission must exist at some level, however,
due to the positrons emitted by nucleosynthesis sources in the disk.
The image also suggests that annihilation near localized sources 
(in particular microquasars, pulsars, or individually-bright supernovae)
does not dominate the e$^+$ budget in the inner Galaxy, and rather a large
number of sources distributed over a larger region, or distributed source
processes, provide the origin of positrons.

\section{Summary and Outlook}
Measurement of characteristic gamma-rays from radioactive isotopes
provides a useful complement to other means of the study of cosmic
nucleosynthesis. Ge spectrometers have been deployed in space,
capable to measure details of characteristic gamma-ray lines
with sufficient precision to directly constrain abundances and kinematics
of freshly-produced radio-isotopes and positrons. 
The diffuse \Al line has been found to be rather narrow. 511~keV emission
appears dominated by an extended source region in the inner Galaxy; dust
plays a significant role in the annihilation process. Neither new \Ti \cite{rena04} nor \Na \cite{jean04}
sources have been found; analysis for \Ti from Cas~A and SN1987A is in progress. 
INTEGRAL is scheduled to observe the gamma-ray sky for at least six years, promising
to significantly advance our understanding of supernovae and novae, and
of the interstellar medium embedding the sources of nucleosynthesis.

\end{document}